\documentclass[aps,prl,twocolumn,superscriptaddress]{revtex4-1}

\usepackage{hyperref} 
\usepackage{graphicx}
\usepackage{bm}
\usepackage{amsmath}
\usepackage{amssymb}

\DeclareGraphicsExtensions{.png,.jpg,.eps}
\usepackage{xcolor}

\begin{document}

\author{Wei Chen}
\affiliation{Institute for Theoretical Physics, ETH Zurich, 8093 Zurich, Switzerland}
\affiliation{College of Science, Nanjing University of Aeronautics and Astronautics, Nanjing 210016, China}

\author{J. L. Lado}
\affiliation{Institute for Theoretical Physics, ETH Zurich, 8093 Zurich, Switzerland}

\title{Interaction driven surface Chern insulator in nodal line semimetals}

\begin{abstract}
Nodal line semimetals are characterized by nontrivial bulk-band crossings,
giving rise to almost flat drumhead-like surface states (DSS),
which provide an attractive
playground where interaction can induce
symmetry-broken states and potential emergent phases.
Here, we show that electronic interaction
drives a Stoner ferromagnetic instability in the DSS
while the bulk remains non-magnetic, which together with
spin-orbit coupling drive the surface states into a 2D Chern insulator.
We show that each piece of DSS carries a half-integer topological charge,
which for systems containing two pieces of DSS
yield a net Chern number $\mathcal{C}=-1$.
We show that this phenomenology
is robust against chiral-symmetry
breaking, that gives a finite dispersion to the DSS.
Our results show that nodal
line semimetals are a promising platform to implement
surface Chern insulators and dissipation-less electron transport
by exploiting enhanced interaction effects
of the DSS.
\end{abstract}

\date{\today}

\maketitle

Topological electronic states
have motivated large research efforts due to their
gapped bulk coexisting with
protected gapless surface modes
\cite{Hasan10rmp,Qi11rmp,Benalcazar17scn,Schindlereaat18scnadv,Langbehn17prl,Song17prl,Wang17prl}.
In particular, chiral edge states are
especially attractive  as they would yield unidirectional channels lacking
electric loss, representing
a cornerstone in low consumption electronics. Natural
compounds for Chern insulator
have been proven to be rather elusive,
motivating several proposals for its realization \cite{PhysRevB.82.161414,Xiao2011,PhysRevB.92.115417,Liu2016},
yet the most successful implementation
requires a building block that is also very rare in nature:
magnetically doped
topological insulators \cite{Chang13scn,Yu10scn,Liu08prl}.
Thus, a key question is whether if Chern
insulators can be engineered by means of a family of materials
more common in nature, which would open new possibilities
in condensed matter research, apart from applications
in  low consumption electronics.

During the last years, the classification of topological insulators
has been extended to so-called topological
semimetals~\cite{Weng16jpcm,Armitage18rmp}, i.e., systems that are gapless in
the bulk and simultaneously host topologically protected surface states.
The topological band crossing may occur
at discrete points or along closed loops in reciprocal space. The former case
corresponds to Weyl/Dirac
semimetals~\cite{Murakami07njp,Wan11prb,Weng15prx,Xu15scn}, whereas the latter
is referred as nodal line semimetals
(NLSMs)~\cite{Burkov11prb,Kim15prl,Yu15prl,Heikkila11jetp,Xu11prl,Weng15prb,Chen15nl,Zeng15arxiv,Fang15prb,Yamakage16jpsj,Xie15aplm,Chan16prb,Zhao16prb,Bian16prb,Bian16nc, Bzdusek16nat,Yan18np,Chenwei17prb,Yan17prb,Ezawa17prb,Bi17prb,Yan16prl,Yan17prb2,Du17npj,Xu17prb,Huang16prb,Zhang17prb}.
The nodal line carries a $\pi$ Berry
flux~\cite{Burkov11prb}, resulting in drumhead-like surface states (DSS)
\cite{Weng15prb}. In the presence of chiral symmetry, such DSS are perfectly flat,
so that
any residual electronic interaction would overcome the surface kinetic energy,
providing a perfect platform to realize strongly correlated and symmetry-broken surface states \cite{Kopnin11prb,Liu17prb,Pamuk17prb,Lothman17prb,Roy17prb}. Very recently,
the spontaneous magnetization in the flat band of zigzag graphene nanoribbon
was observed \cite{Slota18nat,Fujita96jpsj,Wakabayashi99prb,Son06nat},
indicating that the same physics
may exist in the DSS,
a higher-dimensional analogy of
the 1D flat band in graphene nanoribbon \cite{Fang16cpb}.

\begin{figure}[t!]
 \centering
                \includegraphics[width=\columnwidth]{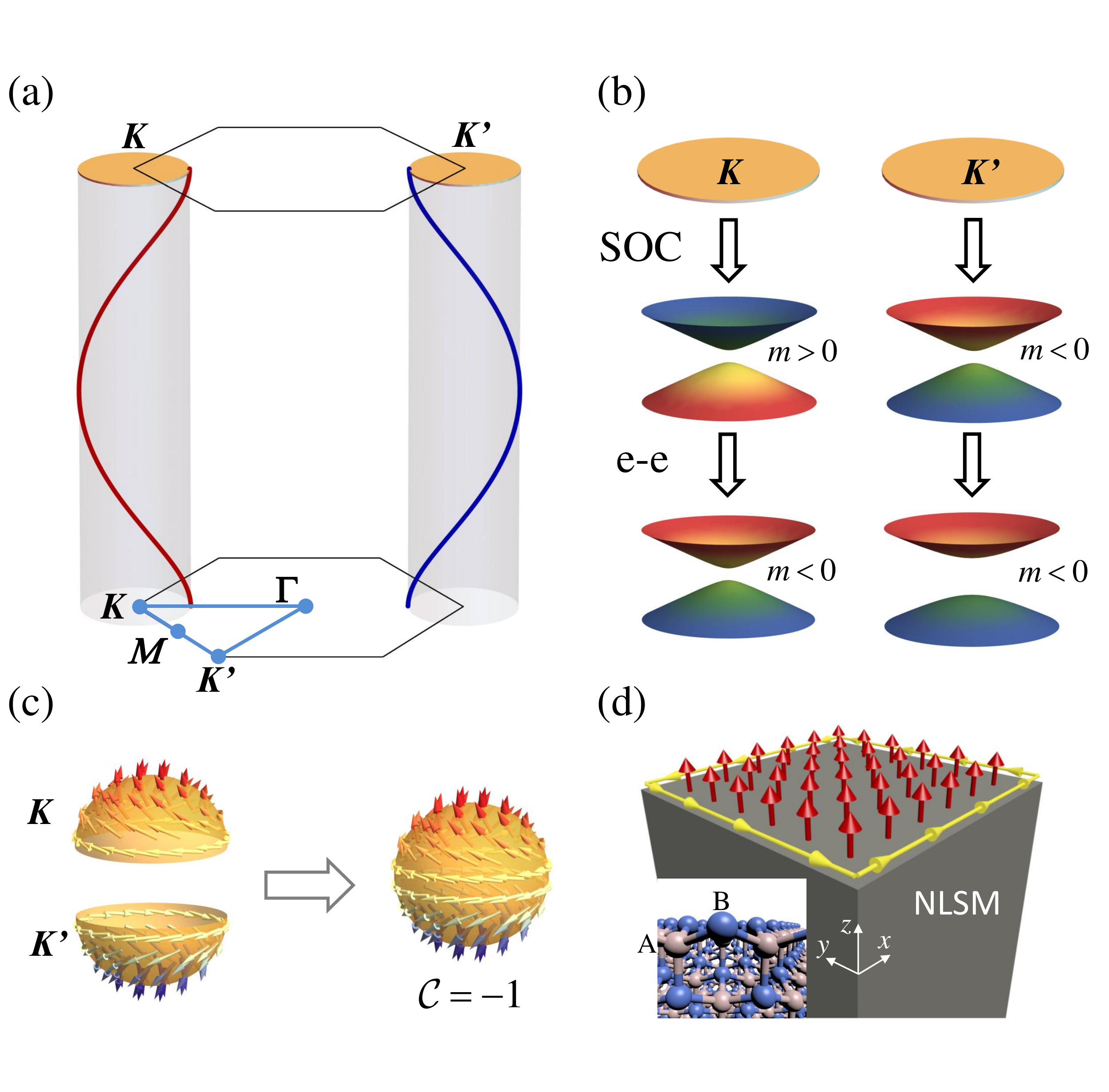}

\caption{(a) Nodal lines around $\bm{K}, \bm{K'}$ points and corresponding DSS (light-orange disks)
enclosed by their projection onto the surface Brillouin zone. (b) Spin-orbit
coupling (SOC) introduces opposite mass terms to the DSS around $\bm{K},
\bm{K'}$. Electronic interaction (e-e) inverts one of the bands. (c) Two pieces of DSS each
carries a meron spin texture with a half-integer topological charge and together
constitute a skyrmion and result in a Chern number $\mathcal{C}=-1$. (d) Spin polarized surface states and chiral hinge
current. Inset: diamond lattice with an open surface.
}
\label{fig1}
\end{figure}

In this Letter, we show that a surface 2D Chern insulator
can emerge from electronic interaction in a NLSM.
For the sake of concreteness,
we focus on the NLSM containing two disconnected pieces
of DSS with spin degeneracy, see Fig. \ref{fig1}(a).
Electronic interaction results in Stoner instability
in the DSS and induces a surface ferromagnetic order.
The emergent exchange field together with the
spin-orbit coupling (SOC) on an open surface, yield a half-integer
topological charge for each piece of DSS, driving
the surface band into a Chern insulator, see Figs. \ref{fig1}(b,c).
Such spontaneously emergent 2D Chern insulator on the
surface of a 3D sample would manifest in 1D chiral hinge
states, see Fig. \ref{fig1}(d). Recent
experimental progress on
fabrication and detection of
the ferromagnetic surface states of topological
insulators paves the way to the realization of our
scheme by using state-of-the-art
techniques \cite{Chen10scn,Wray11np,Mogi17sa,Xiao18prl}.
Our proposal highlights that NLSMs
hosting two nodal lines
that develop a trivial
gap with SOC~\cite{Fang16cpb}
are ideal candidates to realize
a Chern insulator.
In particular, spinel compounds with chemical
composition $XY_2Z_4$ are ideal candidates in this
line, as the $X$ sites form a diamond lattice,
a paradigmatic example of a system hosting two nodal lines \cite{supmat}.

The surface Chern insulator outlined above can be constructively
derived by a concrete model
of NLSM on a diamond lattice,
whose Hamiltonian is
\begin{equation}\label{H}
H = H_0 + H_{SOC} + H_{U}
\end{equation}
where $H_0$ captures the NLSM, $H_{SOC}$ is the
intrinsic SOC and $H_{U}$ is the local
Coulomb interaction term, that we will discuss in detail below.
The Bloch Hamiltonian for the NLSM is
$
H_0=d_x(\bm{k})\sigma_x+d_y(\bm{k})\sigma_y$ with $
d_x(\bm{k})=t+t'\cos(\bm{k}\cdot\bm{a}_3)+t\sum_{i=1}^2\cos(\bm{k}\cdot\bm{a}_i)
$ and $
d_y(\bm{k})=t'\sin(\bm{k}\cdot\bm{a}_3)+t\sum_{i=1}^2\sin(\bm{k}\cdot\bm{a}_i),
$
where Pauli matrices $\sigma_{x,y}$ act on the AB sublattice space [inset in
Fig. \ref{fig1}(d)], $t,t'$ are the nearest-neighbor hopping, and $t'$
denotes the hopping along the $(1,1,1)$ orientation, which has been
set to the $z$-axes for simplicity. The corresponding lattice vectors are
$\bm{a}_1=a(\frac{1}{2\sqrt{2}},\frac{\sqrt{3}}{2\sqrt{2}},0),
\bm{a}_2=a(-\frac{1}{2\sqrt{2}},\frac{\sqrt{3}}{2\sqrt{2}},0),\bm{a}_3=a(0,\frac{\sqrt{6}}{6},\frac{\sqrt{3}}{3})$,
with $a$ being the lattice constant. When $t'/t<1$, the system is a
NLSM, carrying two spiral nodal lines. In a thick
two-dimensional slab, the previous nodal lines
get projected around
$\bm{K}=(-\frac{4\sqrt{2}}{3a}\pi,0)$ and $\bm{K'}=-\bm{K}$ in the $k_x$-$k_y$
plane of the
two-dimensional Brillouin zone\cite{Mcclure69,Heikkila11jetp,Hyart2018,lado2018two}, see Fig. \ref{fig1}(a). This can be seen in the limiting case $t'\rightarrow
0$ by expanding $H_0$ around $\pm\bm{K}$ points as $h^\pm=d_x^\pm\sigma_x+d_y^\pm\sigma_y$, with $d_x^\pm=\pm
vq_x+t'\cos(k_za/\sqrt{3})$ and $d_y^\pm=-vq_y+t'\sin(k_za/\sqrt{3})$, where
the velocity is defined as $v=\sqrt{3}ta/(2\sqrt{2})$, the small wave vector
$\bm{q}=(q_x, q_y)$ is measured from $\pm\bm{K}$. By
putting $d_{x,y}^\pm=0$, we obtain the parametric equations of the spiral nodal
lines around $\pm\bm{K}$ as $q_x=\mp\frac{t'}{v}\cos \frac{k_za}{\sqrt{3}},
q_y=\frac{t'}{v}\sin \frac{k_za}{\sqrt{3}}$, which have opposite chirality
[Fig. \ref{fig1}(a)].

\begin{figure}[t!]
 \centering
                \includegraphics[width=\columnwidth]{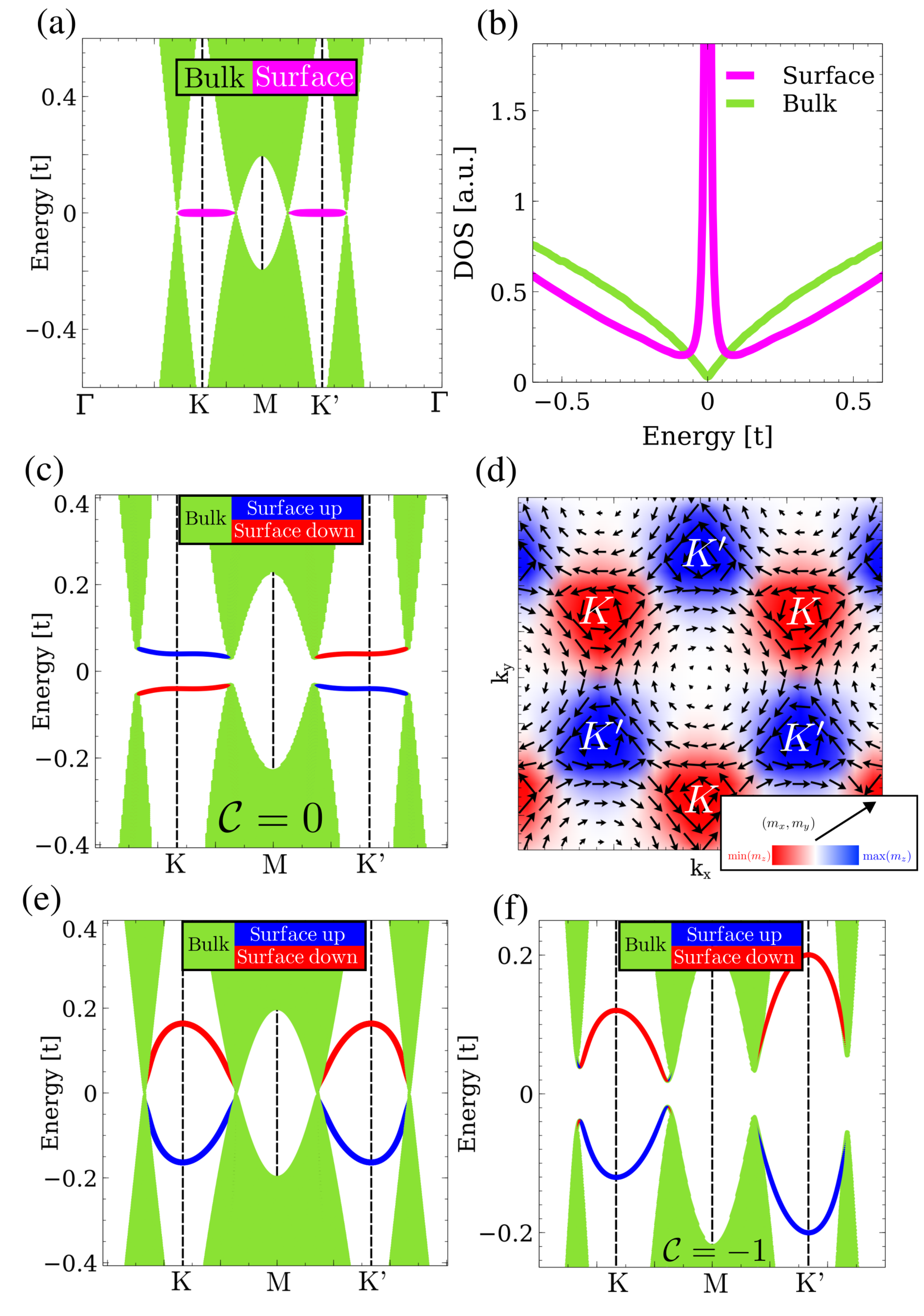}

\caption{(a) Band
structure of a slab of a NLSM showing the surface flat bands. (b) Density of
states (DOS) in the bulk and on the surface. (c) Band structure with only SOC. (d)
Spin texture of the surface states. (e) Band structure with only interaction
effect. (f) Band structure with both SOC and interaction effects.
The slab consists of 600 layers and we
took $t'=0.8t$, $\lambda=0.01t$ and $U=2t$.
}
\label{fig2}
\end{figure}

The nontrivial band topology of the NLSM is characterized
by $\pi$ Berry phase
carried by each nodal line. The Hamiltonian $H_0$ possesses time reversal ($T$) symmetry,
$H_0(\bm{k})=TH_0(\bm{k})T^{-1}=H_0^*(-\bm{k})$, and inversion ($P$) symmetry,
$H_0(\bm{k})=PH_0(\bm{k})P^{-1}=\sigma_xH_0(-\bm{k})\sigma_x$, so that the
Berry curvature vanishes everywhere
away from the nodal lines \cite{Fu07prb}, and the nodal lines are protected by
the combined $PT$ symmetry \cite{Kim15prl}.
Then one can choose an arbitrary
integral path to calculate the Berry phase.
Here we choose the integral path to be a straight line along the
$z$-direction, then the Zak phase calculated inside/outside the projection of
the nodal line equals $\pi/0$. This configuration of line integral is
convenient to show the bulk-boundary correspondence of the NLSM, that is, DSS
appear inside the projection of the nodal lines onto the surface Brillouin zone
[Fig. \ref{fig1}(a)]. For a semi-infinite $(z<0)$ sample with an open
surface lying at $z=0$ [Fig. \ref{fig1}(d)], by substituting $k_z\rightarrow
-i\partial_z$ in $h^\pm(\bm{q})$, the zero-energy DSS around $\bm{K}, \bm{K'}$
can be obtained as \cite{Heikkila11jetp}
$\psi_\pm\propto(0,1)^{\text{T}}e^{\lambda_\pm z}$, where
$\lambda_\pm=\frac{\sqrt{3}}{a}(\ln\frac{t'}{v|\bm{q}|}+i\theta_\pm)$, with
$\theta_-=\tan^{-1}(q_y/q_x), \theta_+=\pi-\theta_-$. As long as
$v|\bm{q}|<t'$, i.e., the states lie inside the projection of the nodal loops,
the wave functions $\psi_\pm$ decays to zero as $z\rightarrow-\infty$,
indicating the existence of
sublattice polarized
DSS [cf. \ref{fig1}(a)]; Otherwise, there are no surface states.
The above
results hold generally in the NLSM regime $t'/t<1$. This
can be checked by computing the band
structure in a slab with
the Hamiltonian $H_0$, infinite
in the $x$-$y$ plane and whose $z$-axis lies
along the $(1,1,1)$ direction of the parent diamond lattice
[inset of Fig. \ref{fig1}(d)].
We take a slab thick enough so that
the two surfaces are decoupled,
and we project the final result onto
the upper half of the system
to retain only the DSS on the upper surface.
We show in Fig. \ref{fig2}(a)
the band structure of the NLSM described above, where
two pieces of zero-energy DSS exist.
Without dispersion, the surface density of states (DOS) diverges at
zero energy, see Fig. \ref{fig2}(b), whereas the bulk DOS vanishes.

Next, we include the SOC effect by second-neighbor hopping \cite{Fu07prl}
as
$H_{SOC}=
i \lambda\sum_{\langle\langle i,j \rangle\rangle}c_i^\dag\bm{s}\cdot(\bm{d}_{jk}\times\bm{d}_{ki})c_j$,
where $c_i=(c_{i\uparrow}, c_{i\downarrow})$ is the Fermi operator for both
spins on site $i$, $\lambda$ is the SOC strength, $\bm{s}$ is the spin vector,
and $\bm{d}_{ik}$ is the vector connecting
sites $i$ and $k$, and $k$ is an intermediate site
between $i$ and $j$. The SOC term opens a gap in the band structure,
both in the bulk and in the surface modes,
lifting the spin degeneracy of the DSS [Fig. \ref{fig2}(c)]
and introducing
nontrivial spin textures to the DSS [Fig. \ref{fig2}(d)].
Its effect on the surface modes
can be described by the following massive Dirac Hamiltonian
$
\mathcal{H}^\pm_{SOC}=\alpha(s_xq_y-s_yq_x)\pm\beta s_z,
$
with the superscript $``\pm"$
corresponding to $\bm{K}$ and $\bm{K}'$ points, which is sufficient to characterize the surface
band topology~\cite{supmat}. Here,
$\mathcal{H}^\pm_{SOC}$ is induced by the bulk SOC on an open surface, so that
two coefficients $\alpha,\beta (>0)$ are determined by $\lambda$~\cite{supmat}.
The parameter $\beta$ introduces opposite mass
terms to the DSS around $\bm{K}, \bm{K'}$ points, see Fig. \ref{fig1}(b).
Without SOC, the Hamiltonian for the DSS vanishes, so that
$\mathcal{H}^\pm_{\text{SOC}}$ can serve as the effective Hamiltonian of the DSS. The Berry
flux carried by each piece of DSS is calculated through
$Q_\pm=\frac{1}{4\pi}\int\int
dq_xdq_y(\partial_{q_x}\bm{b}^\pm\times\partial_{q_y}\bm{b}^\pm)\cdot\bm{b}^\pm/|\bm{b}^\pm|^3$,
with $\bm{b}^\pm=(\alpha q_y,-\alpha q_x,\pm\beta)$, yielding a
topological charge, or meron number \cite{Bernevig2013} $Q_\pm=\pm\frac{1}{2}$~
\footnote{The meron number
will be half-quantized only in the low energy limit, when the mass
term is much smaller than the non-linear corrections
to the dispersion, so that the Berry curvature is heavily
concentrated around $\bm{K},\bm{K}'$ points.}.
Each piece of DSS carries a meron number, but with an opposite sign, due to the
opposite mass term. Two meron numbers thus
cancel out and result in zero Chern
number as imposed by time reversal symmetry.
In this scenario, it is suggestive to think that, if one of the
meron number would be inverted, the
system would show a net Chern number [Fig. \ref{fig1}(c)].

We now show that electronic interactions can spontaneously
break time reversal symmetry
on the surface, inverting one of the meron number and turning the
DSS into a Chern insulator.
For that goal, it is convenient to first consider the case without SOC,
where the system shows gapless flat DSS.
In this situation, the infinite surface DOS would yield a Stoner ferromagnetic instability by
arbitrarily
small interaction, spontaneously breaking the time reversal symmetry on the
surface. In contrast, for small $U$,
no symmetry breaking
occurs in the bulk states.
The surface symmetry breaking can be captured
by adding an interaction term to the single particle Hamiltonian
of the Hubbard form $H_U=U\sum_i
n_{i\uparrow}n_{i\downarrow}$, with
$n_{i\uparrow,\downarrow}=c^\dag_{i\uparrow,\downarrow}c_{i\uparrow,\downarrow}$ the number operator.
Taking
magnetization along the $z$-direction,
in the mean-field picture
the Hubbard interaction can be decoupled as
$H_U \approx H_{MF} =
U\sum_{i} [
n_{i\uparrow} \langle n_{i\downarrow} \rangle +
n_{i\downarrow} \langle n_{i\uparrow} \rangle -
\langle n_{i\uparrow} \rangle \langle n_{i\downarrow} \rangle ]
$.
The magnetization on site $i$ is defined as
$m^z_i=\langle n_{i\uparrow}\rangle - \langle n_{i\downarrow}\rangle$.
The band structure of the self-consistent solution is shown in Fig.
\ref{fig2}(e), with spin-polarized surface states and unpolarized bulk modes.
In terms of the low energy model, electronic
interaction results in an effective Zeeman term
to the Hamiltonian for the DSS. In particular, at the $\bm{K}, \bm{K}'$
points, the new term takes the form
$
\mathcal{H}_{Z}=-m_Z
s_z
$,
where its strength can be evaluated as $m_Z=\frac{U}{2}\int_{-\infty}^0
m^z(z)|\psi_\pm(\bm{q}=0,z)|^2dz$.

Finally, we consider the simultaneous action of both electronic interaction
and SOC. By numerically solving
the full self-consistent model with SOC, we observe that the
surface magnetization survives even in the
presence of SOC, see Fig. \ref{fig2}(f).
In this situation, the effective Hamiltonian
for the DSS takes the form
$\mathcal{H}=\mathcal{H}_{SOC}^\pm+\mathcal{H}_Z$.
Now the Chern number for the whole surface bands can be defined by the
mass terms at the $\bm{K}, \bm{K}'$ points as
\begin{equation}
\mathcal{C}=\frac{1}{2}\big[\text{sgn}(\beta-m_Z)+\text{sgn}(-\beta-m_Z)\big].
\end{equation}
As $m_Z>\beta$,
the surface Zeeman splitting reverses the sign
of the mass term around $\bm{K}$ [compare Figs. \ref{fig2}(c), \ref{fig2}(f)], and
drive the system to a Chern insulator [cf. Fig. \ref{fig1}(b)].
Such kind of topological phase transition resembles
the scenario of the Haldane model on the graphene lattice \cite{Haldane88prl},
yet here the sign change of the mass around one valley
is dynamically generated by electronic interaction.

\begin{figure}[t!]
 \centering
                \includegraphics[width=\columnwidth]{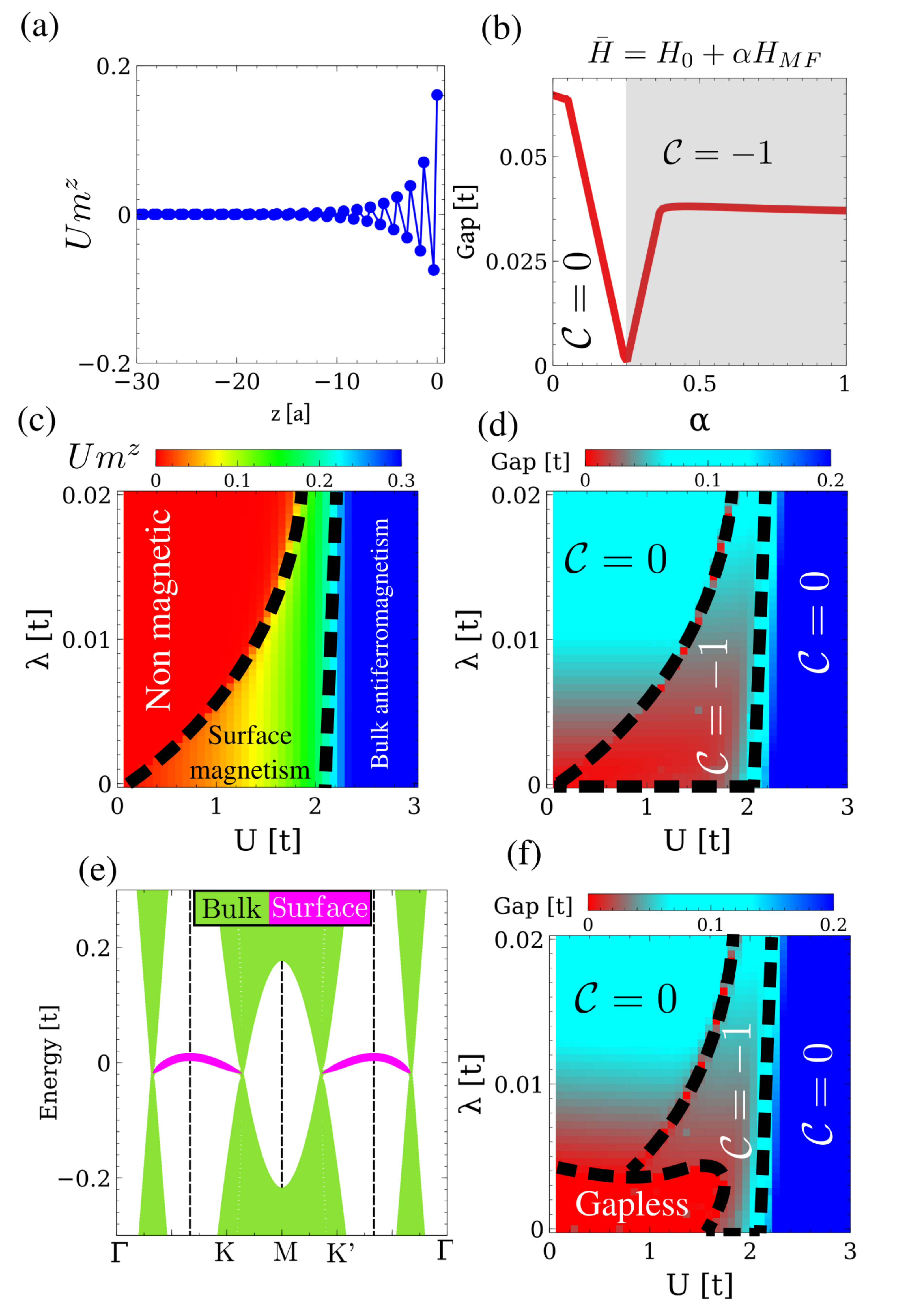}

\caption{(a) Real space distribution of the magnetization,
where $z=0$ corresponds to the surface. (b) Energy gap as a
function of the mean-field order parameter. Phase diagram associated with (c)
surface magnetization and (d) energy gap in terms of SOC strength $\lambda$ and
interaction $U$. (e) Band structure including the second-neighbor hopping $t_2=0.03t$, in the
absence of $U$ and $\lambda$. (f) Phase diagram containing chiral-symmetry breaking. We took $t'=0.8t$, $\lambda=0.01t$ and $U=2t$ for (a,b),
200 layers for (a,b,c,d,f) and 600 layers for (e).
}
\label{fig3}
\end{figure}

Since we are solving a self-consistent problem
that does not have a smooth behavior, a gap
closing and reopening cannot generically
be observed.
Nevertheless,
since the mean-field term of the original Hamiltonian
effectively reduces to a site-dependent
exchange field that decays
as one enters the bulk [Fig. \ref{fig3}(a)],
we may try to artificially switch on its contribution,
in order to adiabatically trace the topological
phase transition.
This can be made concrete by taking
a final self-consistent
Hamiltonian realizing the Chern insulating state
$H = H_0 + H_{MF}$,
and defining an adiabatic Hamiltonian of the form
$ \bar H(\alpha) = H_0 + \alpha H_{MF}$, where
$\alpha=0$ corresponds to the
non-interacting Hamiltonian with $\mathcal{C}=0$,
whereas $\alpha=1$ corresponds to
the physical self-consistent solution with $\mathcal{C}=-1$.
By tuning $\alpha$ from 0 to 1,
the topological phase transition can be observed
[Fig. \ref{fig3}(b)]
as
a gap closing and reopening in the energy spectra, concomitant with a change of
Chern number from 0 to -1.

Since SOC opens up a gap
in the single particle spectra, it is expected
that at large values of $\lambda$ the
magnetic order
will be quenched and the system will remain
a trivial semiconductor.
This competition
between SOC and electronic
interaction is shown in the
phase diagram in terms of $U$ and
$\lambda$ in Figs. \ref{fig3}(c,d). Different from the band closing and
reopening by continuously tuning the order parameter in Fig. \ref{fig3}(b), the
surface magnetization [Fig. \ref{fig3}(c)] and the energy gap [Fig.
\ref{fig3}(d)] change abruptly with varying interaction $U$, indicating a phase
transition with spontaneous symmetry breaking. Remarkably, such a conventional
phase transition further induces and coincides with a topological phase
transition on the surface.
The topologically nontrivial phase with $\mathcal{C}=-1$ holds in a wide
parametric region. For a larger $\lambda$, the parametric region of $U$ for
a Chern insulator becomes narrower, while the energy gap increases.
For large
$U$, the whole system becomes a trivial
anti-ferromagnetic insulator, opening a large bulk magnetic gap.

\begin{figure}[t!]
 \centering
                \includegraphics[width=\columnwidth]{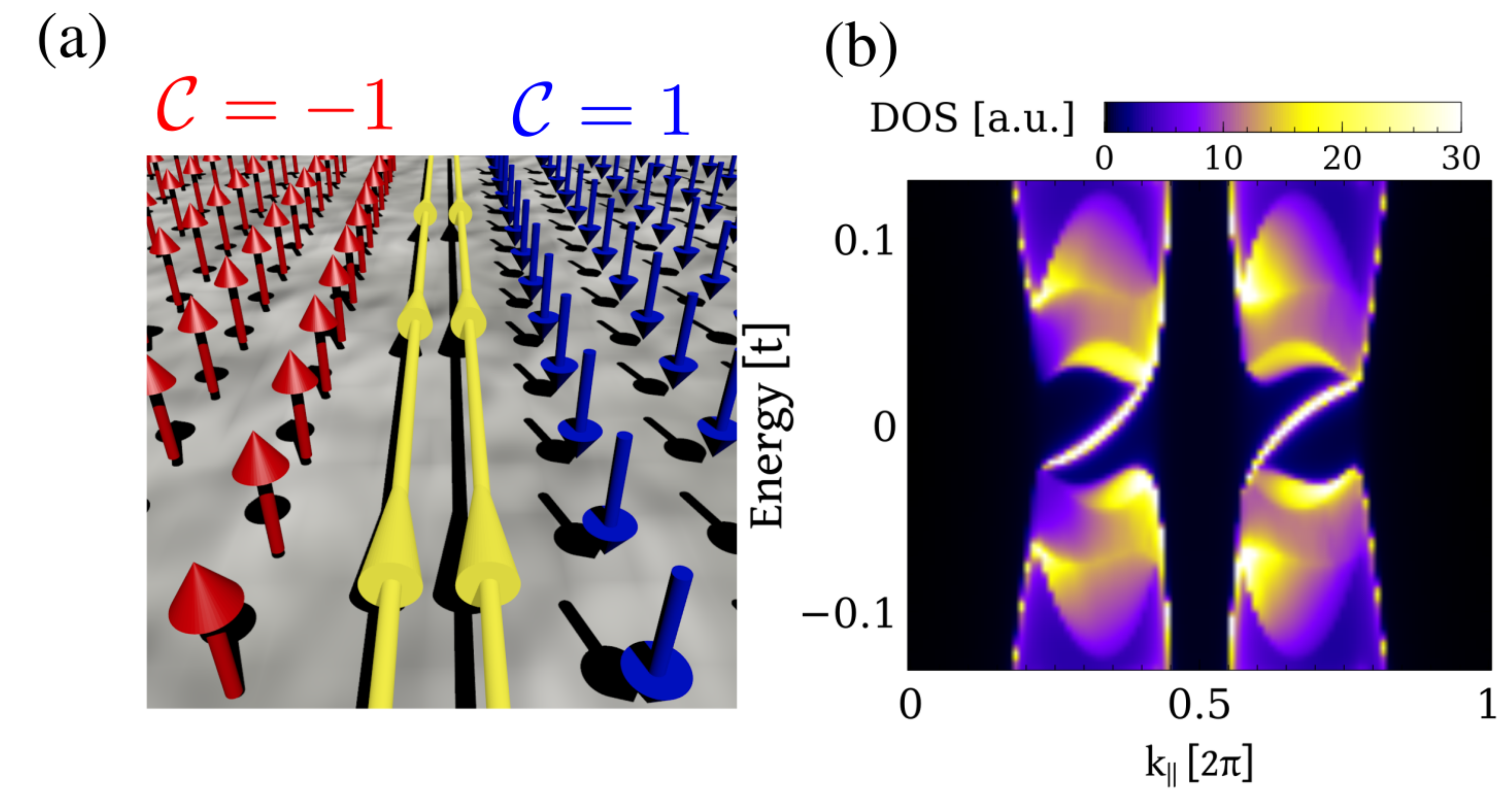}

\caption{(a) Sketch of the interfacial
states in a magnetic domain wall
on the surface of the NLSM.
(b) Surface spectral function of the magnetic domain
wall depicted in (a), showing two copropagating
states along the domain wall.
}
\label{fig4}
\end{figure}

The NLSM described by $H_0$ in Eq. \eqref{H} possesses chiral
symmetry, $\sigma_zH_0\sigma_z=-H_0$, which leads to flat DSS \cite{Hyart2018,Ryu02prl}.
However, in real materials, chiral symmetry is usually broken, and the DSS show a finite
dispersion. We investigate this situation by introducing a second-neighbor
hopping to $H_0$. Dispersive DSS can be seen in the band structure in Fig.
\ref{fig3}(e). The corresponding phase diagram in Fig. \ref{fig3}(f) shows a gapless region, yielding
critical
$U$ and $\lambda$ values for the onset of the surface
Chern insulator. Apart from this,
there is no much difference from the case of flat DSS.

A direct result of a 2D Chern insulator is the existence
of chiral edge states.
Here, the Chern insulator emerges on the surface of a
3D sample, resulting in chiral hinge states [Fig. \ref{fig1}(d)]
\cite{Sitte12prl,Zhang13prl,Benalcazar17scn,Schindlereaat18scnadv,Langbehn17prl,Song17prl,Wang17prl,Li17prb}.
The chiral edge states can also be achieved in a magnetic
domain wall between two oppositely ordered ferromagnetic regions on the
surface. The two
regions are related by time reversal symmetry,
so that they carry opposite Chern numbers, yielding a pair of chiral modes
inside the domain wall, see Fig. \ref{fig4}(a).
This image can be made concrete by computing
the spectral function
of the magnetic domain wall by means of the Dyson equation
$G_D(\omega,k_\parallel) = [\omega - H_D(k_\parallel) -
\Sigma_L(\omega,k_\parallel)- \Sigma_R(\omega,k_\parallel)]^{-1}$,
where $k_\parallel$ is the Bloch momentum
along the direction defined by the domain wall,
$\Sigma_{L,R}(\omega,k_\parallel)$ are the self-energies induced by the
semi-infinite magnetic regions
(taken
from the self-consistent solution of the infinite problem), and
$H_D(k_\parallel)$ is the local Hamiltonian of the
domain wall. With the previous Green's function, we compute
the DOS at the surface as
$\frac{1}{\pi}\text{Im}[\text{Tr}_U(G_D)]$,
where $\text{Tr}_U$ traces over the degrees of freedom of the upper surface.
The interfacial spectral function
is shown in Fig. \ref{fig4}(b),
where it is seen that two gapless modes appear
at the magnetic domain wall.
Therefore, controlling such magnetic
domain walls \cite{Lahtinen2012,McGilly2015}
would allow
to imprint chiral states on the surface of NLSMs.

To conclude, we have demonstrated that correlation
effects and SOC can
drive the NLSM into a surface Chern insulator.
Spin-degenerate $PT$ symmetric NLSMs with two
pieces of DSS in the surface Brillouin
zone are potential candidates to achieve such a topological phase~\cite{supmat}.
For the NLSM in the presence of SOC,
since the nodal line is
robust against SOC, other mechanisms are required to
open a trivial gap in the bulk states~\cite{supmat}.

\begin{acknowledgments}
We thank M. H. Fischer, Qiang-Hua Wang, Oded Zilberberg, M. Sigrist
and D. Y. Xing for
helpful discussions, and Zhong Wang for valuable feedback on the manuscript.
W.~C. acknowledges the support from the National Natural Science Foundation of
China under Grants No. 11504171, and the Swiss Government Excellence
Scholarship under the program of China Scholarships Council (No. 201600160112).
J. L. L. acknowledges financial support from the ETH Fellowship program
	and from the Japan Society
for the Promotion of Science Core-to-Core program "Oxide
Superspin" international network..
\end{acknowledgments}


%

\newpage
\onecolumngrid
\renewcommand{\theequation}{S.\arabic{equation}}
\setcounter{equation}{0}
\renewcommand{\thefigure}{S.\arabic{figure}}
\setcounter{figure}{0}

\section{Supplemental Material for \\Interaction driven surface Chern insulator in nodal line semimetals}

\subsection{Derivation of spin-orbit coupling in the surface states}

\begin{figure}[t!]
\centering
                \includegraphics[width=0.5\columnwidth]{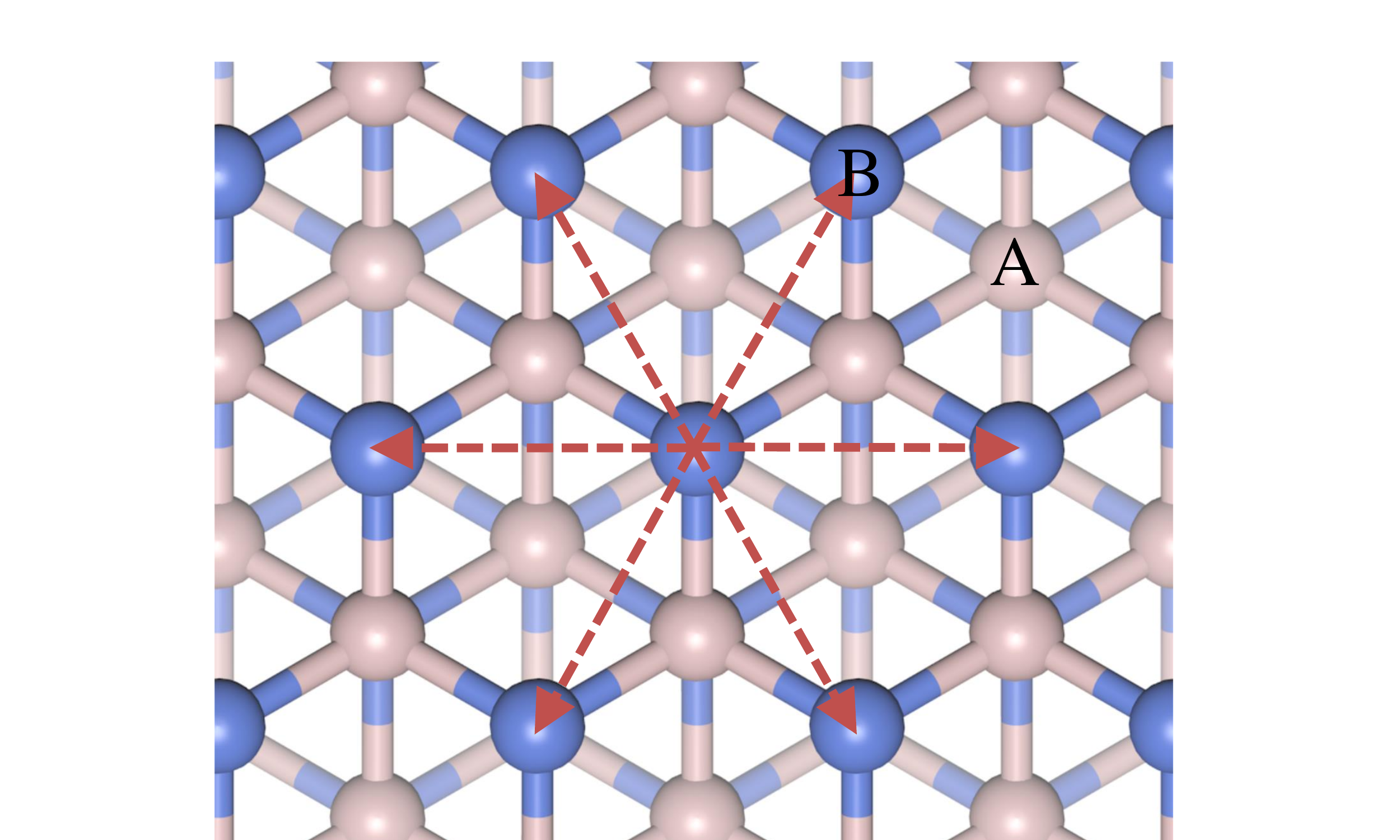}

\caption{Top view of the lattice and the next-nearest-neighbor hopping (dashed arrows) induced
surface spin-orbital coupling.
}
\label{figs1}
\end{figure}

In the following we present an analytic derivation of
the spin texture of the surface states,
shown in Fig. 2(d) of the main manuscript.
The surface spin-orbit coupling (SOC) shows two
important features:
first a vortex-like spin structure of the in-plane component,
and second a net $z$-component in the
center of the nodal line, both
around the $\bm{K}, \bm{K}'$.
To derive those two features, we start with the
bulk SOC Hamiltonian
\begin{equation}
H_{SOC}=i\lambda\sum_{\langle\langle i,j\rangle\rangle}c^\dag_i\bm{s}\cdot(\bm{d}_{jk}\times
\bm{d}_{ki})c_j.
\end{equation}
The surface states
distribute mainly on the top B sublattice, see Fig. \ref{figs1},
so that the surface SOC can be obtained by
retaining the in-plane hopping between nearest B sites
(next-nearest-neighbor hopping for the
whole lattice). Then the surface SOC reduces to
\begin{equation}
\mathcal{H}_{SOC}=i\lambda\sum_{i,\bm{\delta}}c_{i+\bm{\delta}}^\dag\bm{s}\cdot(\bm{d}_{ik}\times\bm{d}_{k,i+\bm{\delta}})c_i,
\end{equation}
where $\bm{\delta}$ denotes the vectors connecting B sites (Fig. \ref{figs1}).
Performing Fourier transformation we obtain
\begin{equation}
\begin{split}
\mathcal{H}_{SOC}(\bm{k})&=i\lambda\sum_{\bm{k}}c^\dag_{\bm{k}}
(b_xs_x+b_ys_y+b_zs_z)c_{\bm{k}}\\
b_x&=-\frac{4\sqrt{2}}{3}\lambda\cos\frac{k_x a}{2\sqrt{2}}\sin\frac{\sqrt{3}k_ya}{2\sqrt{2}}\\
b_y&=\frac{4\sqrt{2}\lambda}{3\sqrt{3}}\Big(2\cos\frac{k_xa}{2\sqrt{2}}+\cos\frac{\sqrt{3}k_ya}{2\sqrt{2}}\Big)\sin\frac{k_xa}{2\sqrt{2}}\\
b_z&=-\frac{16\lambda}{3\sqrt{3}}\Big(-\cos\frac{k_xa}{2\sqrt{2}}+\cos\frac{\sqrt{3}k_ya}{2\sqrt{2}}\Big)\sin\frac{k_xa}{2\sqrt{2}}
\end{split}
\end{equation}

Expanding the Hamiltonian around $\bm{K}$ and $\bm{K}'$ points yields
\begin{equation}\label{soc}
\mathcal{H}_{SOC}^\pm=\alpha(s_x q_y-s_y q_x)\pm\beta s_z,
\end{equation}
where $\alpha=\lambda a/\sqrt{3}$ and $\beta=4\lambda$, and the superscript $``\pm"$
corresponds to $\bm{K}$ and $\bm{K}'$ points, respectively.
Since the open surface breaks
inversion symmetry,
the projection of the bulk SOC
gives rise to a Rashba-type SOC on the surface.

The topological phase
transition occurs when one of the mass term decreases to zero and then changes its sign by the
surface Zeeman splitting. Any smooth modulation of the band structure
without closing the gap will not change the band topology. Thus, we
may adopt a small mass term $\beta\ll 4\lambda$ in Eq. \eqref{soc} to
characterize the topological phase transition. In this limit,
the Berry curvature distributes mainly around $\bm{K}$ and $\bm{K}'$ points,
and the contribution in the other area can be
neglected. Since the band inversion driven by
interaction occurs around $\bm{K},\bm{K}'$ points, the location
of the surface states, it is sufficient to describe the
topological phase transition by the massive Dirac equation of the surface states.

\subsection{Mean field calculations}
Here, we summarize the computational details of the mean field
calculations presented in the main text.
The total Hamiltonian of the system is of the form
$
	H = H_0 + H_U
$
where $H_0$ is the free particle Hamiltonian and $H_U$ is the interaction
Hamiltonian, that has the form

\begin{equation}
H_U=U\sum_i
n_{i\uparrow}n_{i\downarrow}
\end{equation}
	with
	$
n_{i\uparrow,\downarrow}=c^\dag_{i\uparrow,\downarrow}c_{i\uparrow,\downarrow}
$.
In the collinear
mean field approximation, we can decouple the interaction
Hamiltonian into the following form

\begin{equation}
H_U \approx H_{MF} =
U\sum_{i} [
n_{i\uparrow} \langle n_{i\downarrow} \rangle +
n_{i\downarrow} \langle n_{i\uparrow} \rangle -
	\langle n_{i\uparrow} \rangle \langle n_{i\downarrow} \rangle ]
\end{equation}

We now note that the previous Hamiltonian
depends on the expectation values of its ground state, so that
the solution of the full system must be obtained in a self-consistent way.
Note that, at charge neutrality and up to a global shift in energy,
the previous Hamiltonian is equivalent to

\begin{equation}
H_{MF} =-\frac{U}{2}
\sum_{i,\alpha,\beta}
	m^z_i s^{\alpha,\beta}_{z,i} c^\dagger_{\alpha,i} c_{\beta,i}
\end{equation}
with $m^z_i= \langle n_{i\uparrow} \rangle
- \langle n_{i\downarrow} \rangle$, $s_{z,i}$ the third spin Pauli matrix
on site $i$
and $\alpha,\beta$ the spin indexes.
The solution is obtained by determining
a self-consistent interaction-induced exchange field in every site as
$m^z_i=\int n(E_{\bm{k}})
\langle \Psi_{\bm{k}} | s_{z,i} | \Psi_{\bm{k}} \rangle d^2\bm{k}$,
where $| \Psi_{\bm{k}} \rangle$ are the eigenvectors at the Bloch
momenta $\bm{k}$,
and $n(E_{\bm{k}})$ is the occupation number of state with energy
$E_{\bm{k}}$, that follows a Fermi-Dirac distribution. In the present case,
since the system is at half-filling, the Fermi energy will be located at
$E_F=0$, and thus the occupation numbers are
$n(E_{\bm{k}})=1$ for $E_{\bm{k}}<0$
and $n(E_{\bm{k}})=0$ for $E_{\bm{k}}>0$.
The previous equation defines a self-consistent problem with as many parameters
as sites in the unit cell, which can be solved by conventional
iterative procedures.

The collinear mean field approximation is expected to be valid if
the magnetization of the system can be assumed to lie in the $z$-direction.
In the case without spin-orbit coupling, such assumption in clearly valid
because there is no preferred spin direction
and the system is bipartite
and at half filling \cite{PhysRevLett.62.1201}. In the presence
of spin-orbit coupling, anisotropic effects could give rise to a preferred
axis for the magnetization. For example,
for the NLSM
carrying a low-energy model of the
surface states similar to our model,
the surface Rashba spin-orbit coupling may
lead to a small spin canting towards the
in-plane directions~\cite{Liu17prb}. In a real material, this magnetic anisotropy
is expected to depend on details of the surface structure of the material,
and therefore cannot be captured with the present low energy model.
In our calculations we will assume that those additional anisotropic
effects in the drumhead-like surface states give rise to an
off-plane magnetization, in close analogy with the
anisotropic interactions mediated by the surface states of
three dimensional topological insulators~\cite{Chang13scn,
PhysRevLett.102.156603,PhysRevLett.106.136802}.

We now comment on the origin of the ferromagnetic
order of the surface states.
The model we use in our manuscript, a diamond lattice, is bipartite.
In particular, Lieb's theorem~\cite{PhysRevLett.62.1201}
states that for a Hubbard model in a bipartite
lattice, the ground state has a spin
that is equal to the difference between sites
in the two sublattices.
The surface of our system consists only of a single sublattice,
and therefore there is a local sublattice disproportionation,
that following Lieb's theorem will give rise to a net
ferromagnetic moment~\cite{PhysRevLett.62.1201}.
This is the exact same mechanism
as the one responsible of the
ferromagnetic instability in graphene zigzag edges.

\subsection{Remark on nodal line semimetal (NLSM) with SOC}
Nodal lines may appear
in systems with vanishing SOC
and with finite SOC \cite{PhysRevB.92.081201}. Our manuscript
deals with the first type, where the nodal
lines are protected by the $PT$ symmetry in the absence of SOC \cite{PhysRevLett.115.036806}.

In order to obtain a surface Chern insulator,
the bulk states need to be gapped into a trivial insulator.
Otherwise, the surface states cannot be isolated, which
merges with the gapless bulk states,
yielding a net gapless system where the Chern number
cannot be defined. The
advantage of the $PT$ symmetric NLSM is that
the bulk SOC effect can be used to open a trivial gap
in the bulk states. For the NLSM
with SOC, since the nodal line is robust against SOC,
other mechanisms are required to open a trivial gap.

A plethora of NLSMs
are protected by the $PT$ symmetry in the absence of
SOC. In contrast, fewer NLSMs in the presence of SOC are known,
including PbTaSe$_2$ \cite{Bian16nc}, TlTaSe$_2$ \cite{Bian16prb}
and HgCr$_2$Se$_4$~\cite{PhysRevLett.107.186806,Fang16cpb},
which are not good candidates based on our
criterion for the surface Chern insulator listed below.
To achieve a surface Chern insulator in NLSMs in the presence of SOC,
other materials need to
be explored.

Technically, creating a surface Chern insulator
with a system hosting gapless nodal lines
should be possible provided
the following conditions were accomplished:
two (or an even number)  pieces of drumhead-like surface states, surface Rashba-type SOC
induced by breaking inversion symmetry, surface ferromagnetic
order due to the Stoner instability, and a trivial gap opened
in the bulk state. As long as these
conditions are fulfilled,
a surface Chern insulator can also exist in an NLSM with SOC.
First, NLSM in the presence of SOC
with two (or an even number) pieces of drumhead-like
surface states can
be potentially found, as there is no fundamental
obstacle for the existence
of such a semimetal phase.
Second, an open surface naturally break inversion symmetry,
and generally induces a Rashba-type SOC in the surface states,
as in the case of TlTaSe$_2$, where
Rashba-type surface spin texture has been observed \cite{Bian16prb}.
Third, in order to realize surface ferromagnetic order,
the surface states must have a weak dispersion,
and contain the spin degree of freedom. Thus,
the NLSM such as HgCr$_2$Se$_4$ (also known as the
double-Weyl semimetal because of the existence
of two additional nodal points)
\cite{PhysRevLett.107.186806} cannot be a
candidate, because the
drumhead-like surface states
consist of a single spin channel.
Fourth, it should be possible to open up a trivial
gap in the bulk of the system. For $PT$ symmetric
NLSMs this is naturally accomplished by considering the
bulk SOC. For systems that host gapless nodal lines
in the presence of SOC an alternative mechanism is required,
such as breaking the crystal symmetry that protects
the nodal lines, by creating a distortion in the material.

\subsection{Possible candidate materials}
The tight-binding model of the main manuscript
is mathematically
equivalent to the one of rombohedrically stacked graphite \cite{McClure1969},
that has been recently synthesized \cite{Henni2016}.
Therefore, graphite would realize the
mechanism of the manuscript, yet we believe
that our proposal is specially interesting for compounds
involving heavier atoms, so that the topological
surface gap is enhanced.

A family of materials that would be ideally suited for our proposal
are spinels, a family of
materials isostructural to the
ferromagnetic NLSM HgCr$_2$Se$_4$ \cite{PhysRevLett.107.186806}.
In these compounds, whose chemical composition is XY$_2$Z$_4$, the X atoms sit in a
diamond lattice, and thus potentially realize a multi-orbital version
of our model. Interestingly, the spinel family has more than
150 existing materials \cite{Brik2014}, which suggests that
a nonmagnetic compound hosting nodal lines, that
becomes gapped upon introduction of SOC,
is likely to exist within this family.

More generically, our calculations in the continuum limit show
that a similar phenomenology may be expected in generic
NLSMs. Up to this date, there is a plethora
of proposals hosting nodal lines protected by $PT$ symmetry
with vanishing SOC,
in particular in Cu$_3$NPd\cite{PhysRevLett.115.036806},
CaTe\cite{Du2017}, Ca$_3$P\cite{PhysRevB.93.205132},
BaAs$_3$\cite{PhysRevB.95.045136}, BaSn$_2$\cite{PhysRevB.93.201114}
and TiB$_2$\cite{PhysRevB.95.235116}. Turning on SOC
in this class of materials opens up a gap in the
nodal line, in particular in CaAs$_3$, BaAs$_3$ and SrP$_3$ \cite{PhysRevB.95.045136}.
Therefore, these materials are potential
candidates to realize the phenomenology of our model.

\end{document}